\documentstyle[aps,epsf]{revtex}
\def\sech{\hbox{sech}}
\begin{document}
\title{Coherent transport in a diffusive normal wire between two reservoirs}
\author{A.D. Zaikin, F.K. Wilhelm}
\address{Institut f\"{u}r theoretische Festk\"{o}rperphysik,
Universit\"{a}t Karlsruhe (TH), D-76128 Karlsruhe, Germany}
\author{A.A. Golubov}
\address{Institut f\"{u}r Schicht- und Ionentechnik, Forschungszentrum
J\"{u}lich, D-52425 J\"{u}lich, Germany}
\maketitle
\thispagestyle{empty}
\bigskip

{\small We develop a detailed analysis of electron transport in normal
diffusive conductors in the presence of proximity induced
superconductivity. A rich structure of temperature and energy
dependencies for the system conductance, density of states and
related quantities was found and explained. If the normal conductor
forms a loop its conductance changes $h/2e$-periodically 
with the magnetic flux inside the loop. The amplitude of 
these conductance oscillations shows a reentrant behavior and 
decays as $1/T$ at high $T$.}

\bigskip

Presently transport properties of normal-superconducting 
proximity systems attract much experimental and theoretical interest. 
Various nontrivial features of such systems have been recently discovered [1,2]. 
The aim of this paper is to investigate the coherent electron transport in
a normal diffusive conductor attached to a superconductor. 

One can show [3] that complicated geometrical realizations of the system [1,2]
can be essentially reduced to the following simple model: a
normal diffusive wire is attached to a normal reservoir at $x=0$ and a
superconducing one at $x=d$. In order to calculate the conductance of this
wire we use the standard formalism of quasiclassical Green functions in the 
Keldysh technique (see e.g. [4]). The first step is to find the 
retarded normal and anomalous Green functions of the system 
$g^R=\cosh\theta$ and $f^R=\sinh\theta$, $\theta=\theta_1+i\theta_2$.
In the diffusive approximation this has been done with the aid of
the Usadel equation (see [3] for details). The second step is to 
solve the kinetic equation. As a result for a differential conductance 
of the system (normalized to its Drude value $G_N$) at low voltages and in the absence of tunnel barriers one 
finds [4] $\bar{G}={1\over2T}\int_0^\infty
{d\epsilon\;\sech^2(\epsilon/2T)} D(\epsilon)$ where
$D(\epsilon)=\left(\int_0^1d\bar{x}\;{\sech^2\theta_1(\bar{x})}\right)^{-1}$ is the
transparency of the system at the energy $\epsilon$.

The conductance $G(T)$ shows the reentrant behavior (Fig. 1) (see also [5]). At
low temperatures $T\ll\epsilon_d$ the correction is $\delta
G:=\bar{G}-1\propto (T/\epsilon_d)^2$, at $T\gg\epsilon_d$ we have
$\delta G\propto\sqrt{\epsilon_d/T}$.
The
square-root-scaling of $\delta G$ at high $T$ has an obvious physical 
interpretation: as the part of the N-wire of the size 
$\sim \xi_N=\sqrt{D/2\pi T}$ close to the NS boundary becomes
effectively superconducting due to the proximity effect, only the
rest of the wire contributes to the resistance. As a result it becomes
smaller than $1/G_N$. As the temperature is lowered the conductance
increases, reaches its maximum at $\xi_N \sim d$ an then decreases again.

\begin{minipage}[t]{18cm}
  \begin{minipage}[t]{5cm}
  \begin{figure}
    \centerline{\epsfxsize5cm \epsffile{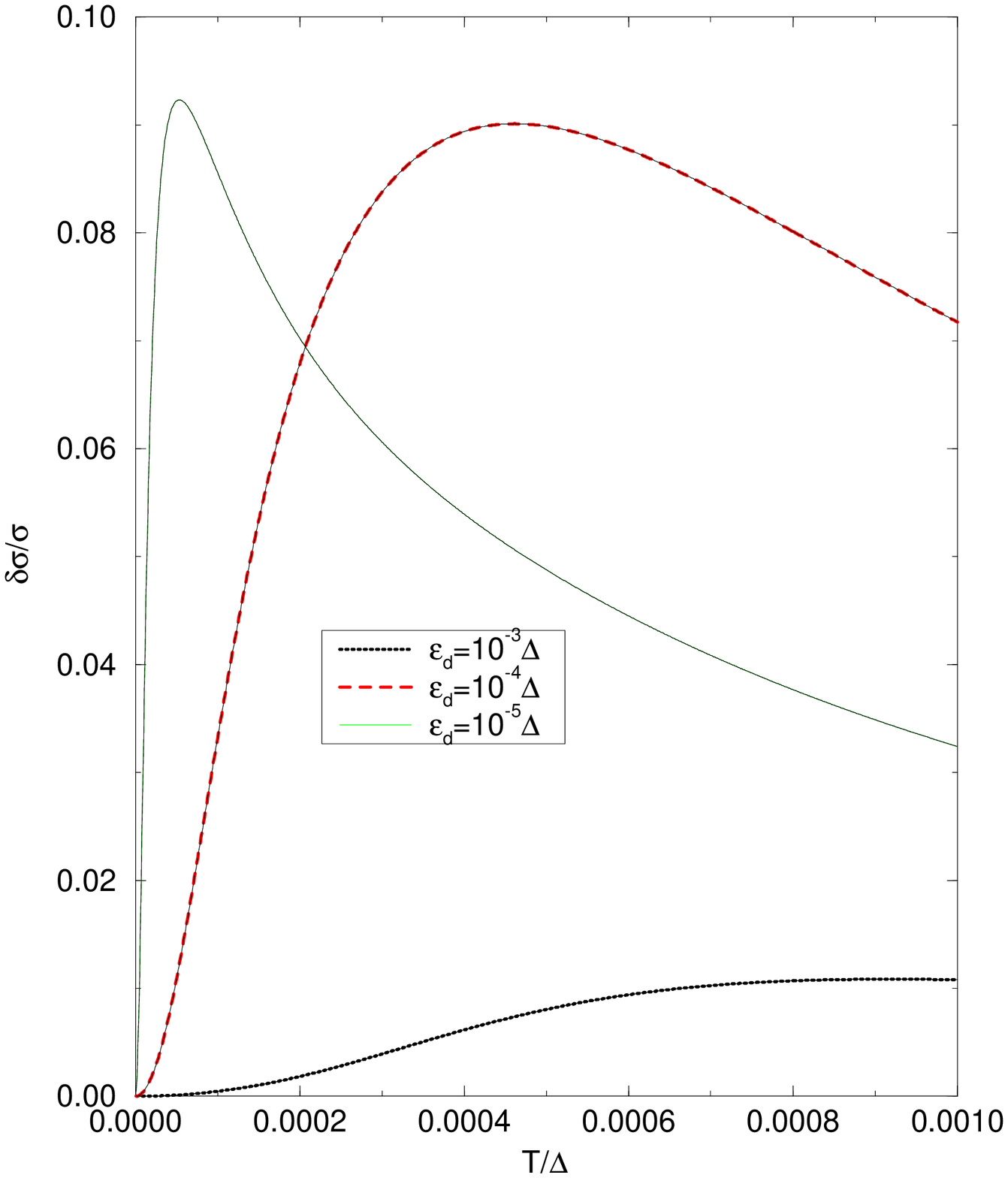}}
  \caption{}
  \end{figure}
  \end{minipage}
  \begin{minipage}[t]{5cm}
    \begin{figure}
    \centerline{\epsfxsize5cm \epsffile{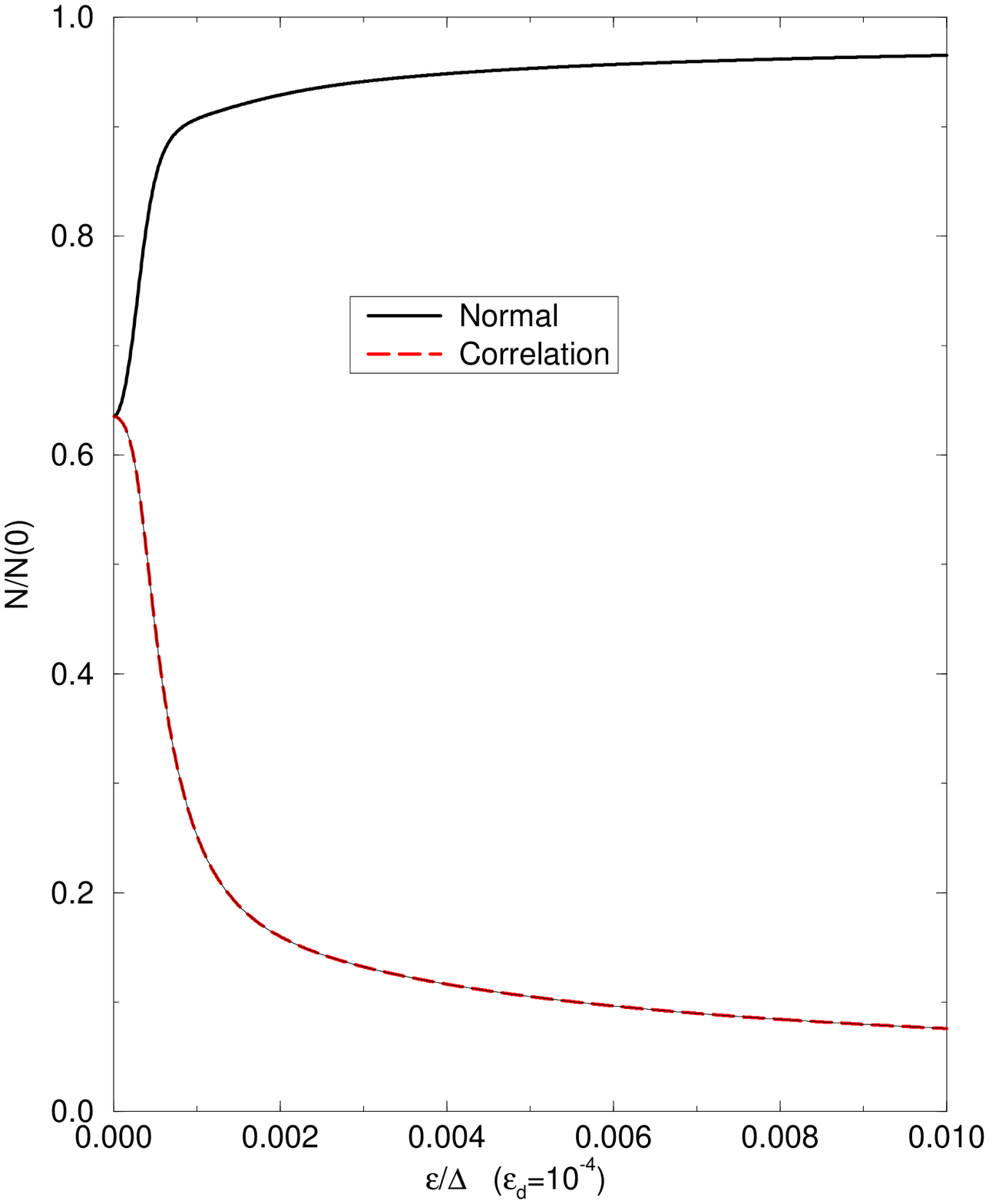}}
    \caption{}
    \end{figure}
  \end{minipage}
  \begin{minipage}[t]{5cm}
    \begin{figure}
    \centerline{\epsfxsize5cm \epsffile{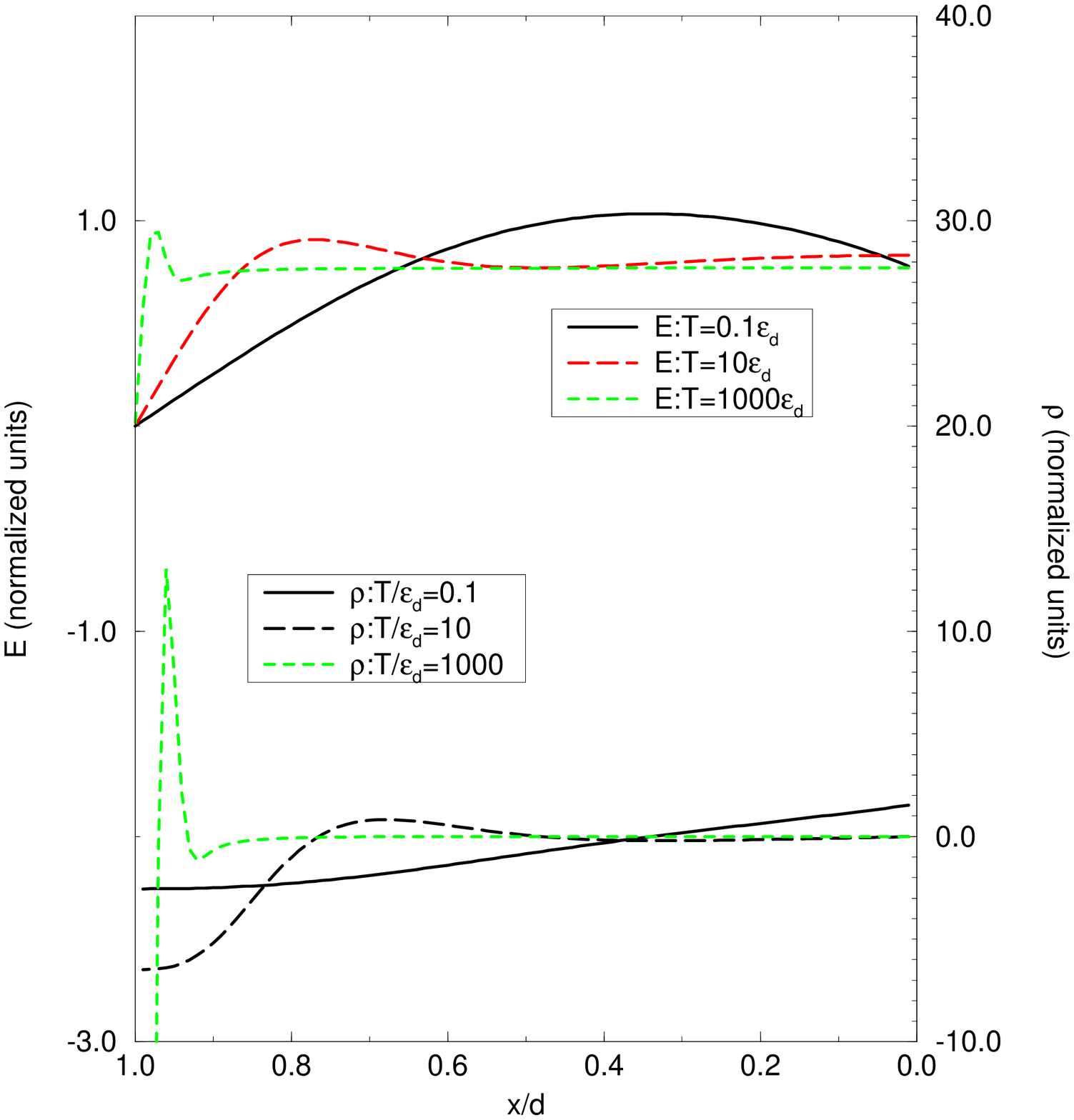}}
    \caption{}
    \end{figure}
  \end{minipage}
\end{minipage}\\

In order to understand the reentrant behavior of $G$ at low $T < \epsilon_d$ 
we calculated the density of states (DOS) averaged over the length of the wire.
The normal density of states $N_N=N(0)\int_0^1d\bar{x}\,\Re(g)$ shows a soft pseudogap below
$\epsilon_d$ (see Fig. 2). At the first sight at low $T$ this would
lead to a decrease
of $\bar{G}$ above 1. This is, however, not the case because of an additional
contribution of correlated electrons present in the N-wire due to the proximity effect.
The DOS for such electrons in the wire $N_S=N(0)\int_0^1d\bar{x}\,\Im(f)$ 
becomes larger for small $\epsilon$ (Fig.2). These two effects exactly 
compensate each other at $T=0$, in which case $\bar{G}=1$. For $T>0$ 
we always have $(\Re g)^2+(\Im f)^2=\cosh^2\theta_1>1$, i.e. the pseudogap effect
never dominates the correlation-induced enhancement and $G(T>0)>1$. On the other
hand, due to the presence of this pseudogap at $\epsilon \sim \epsilon_d$ 
the total transparency $D(\epsilon )$ decreases with $\epsilon$ for 
$\epsilon < \epsilon_d$ resulting in the corresponding reentrant behavior
of $G$ at low $T$. 

Our theory also allows to calculate the profile of the electric field $E(x)$
and the charge $\rho (x)$ inside the N-metal (Fig.3). At $T=0$ we have
(in normalized units) $E(\bar{x})=\cos(\bar{x}\pi/2)-{\pi\over2}(\bar{x}-1)\sin(\bar{x}\pi/2)$
and at $T\gg\epsilon_d$:
$E(\bar{x})\propto(1-\bar{x})\sqrt{T/\epsilon_d}$ near the
superconductor ($1-\bar{x}\ll\xi_T$) and $E(\bar{x})=1$ near the
normal reservoir ($\bar{x}\ll\xi_T$). As we see, the electric field
distribution is essentially nonuniform: it monotoneously decreases 
with $T$ close to a superconductor and overshoots its normal state
value further from it.

Let us also point out that the reentrant behaviour of $G(T)$ takes
place only in the absence of low transparent tunnel barriers at the
boundaries of the N-wire. In the presence of such barriers the field
distribution in the system becomes entirely different and $G(T)$
monotoneously increases with $T$ for most of the situations [3]. Depending 
on the sample both increasing or decreasing $G(T)$ has been found 
in the experiments [1].

In the experiments [2] the conductance of a ring-shaped 
proximity wire was investigated in the presense of the magnetic 
flux $\Phi$ inside the ring. This system shows $h/2e$-periodic 
in $\Phi$ magnetoresistance-oscillations. At high temperatures 
the amplitude of these decays $\propto 1/T$ [2]. 

For simplicity, we have chosen a system where the ring has
circumference $2d$ and the connections to the reservoirs are of the
length $d$, so the Thouless energies $\epsilon_d={\cal{D}\over d^2}$ and
$\epsilon_{3d}={\cal{D}\over (3d)^2}$ become important. We introduced a phase
by a gauge transformation and used a ``Kirchhoff law'' for the Green's
functions at branching points [6,3].

The transparency of the system $D(\epsilon )$ (Fig. 4) shows an interesting
structure. E.g. the oscillations change their sign at
intermediate energies. This effect could possibly be probed by
measurements at very low temperatures ($T\ll\epsilon_{3d}$) at finite
voltages. At high energies ($\epsilon\gg\epsilon_d$), however, the
amplitude of the oscillations of $D(\epsilon)$ is exponentially
supressed because for such energies no superconducting correlations
(which can only originate $h/2e$-oscillations) are present in the ring.

Thus -- even at high temperatures -- only the states with low energies
$\epsilon < \epsilon_d$ are responsible for the effect of conductance 
oscillations. At $T\gg\epsilon_d$ the amplitude of these oscillations
can be estimated as  $\Delta\bar{G}=\bar{G}_{\Phi=0}-\bar{G}_{\Phi=h/4e}
\approx
{1\over2T} \int_0^{\epsilon_c} {d\epsilon}\;\sech^2(\epsilon/2T)(D_0-D_{h/4e})
\approx {\epsilon_c\over2T}\Delta D_{av}$ where 
$\epsilon_c\approx\epsilon_d$ and $\Delta D_{av}$ is the
average amplitude of the oscillations of the transparency below
$\epsilon_c$. This estimate demonstrates that -- in a complete 
agreement with the experimental results [2] --the $1/T$-scaling persists
at all temperatures $T >\epsilon_d$. 
Also the numerical results show that the $1/T$-scaling of
the oscillation amplitude is
excellently fulfilled at sufficiently high $T$. (see Figs. 5 and 6).
At lower temperatures the amplitude of the conductance oscillations 
shows the reentrant behavior reaching the maximum at $T \sim \epsilon_d$
and vanishing completely at $T=0$ (Fig. 5). This reentrant behavior
has a similar physical origin to that of $G(T)$ discussed above in
the absence of the ring.

\begin{minipage}[t]{18cm}
  \begin{minipage}[t]{5cm}
  \begin{figure}
    \centerline{\epsfxsize5cm \epsffile{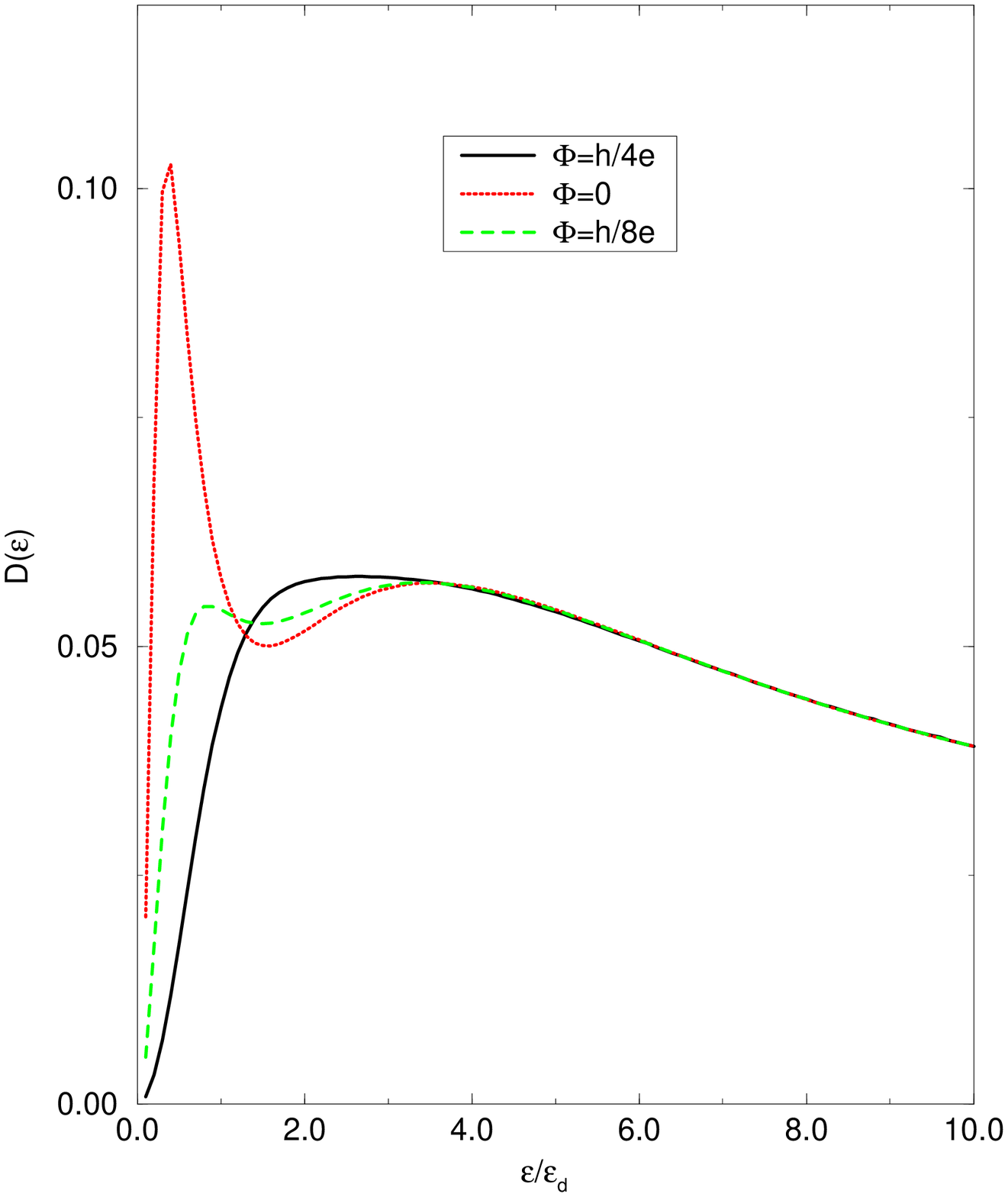}}
  \caption{}
  \end{figure}
  \end{minipage}
  \begin{minipage}[t]{5cm}
    \begin{figure}
    \centerline{\epsfxsize5cm \epsffile{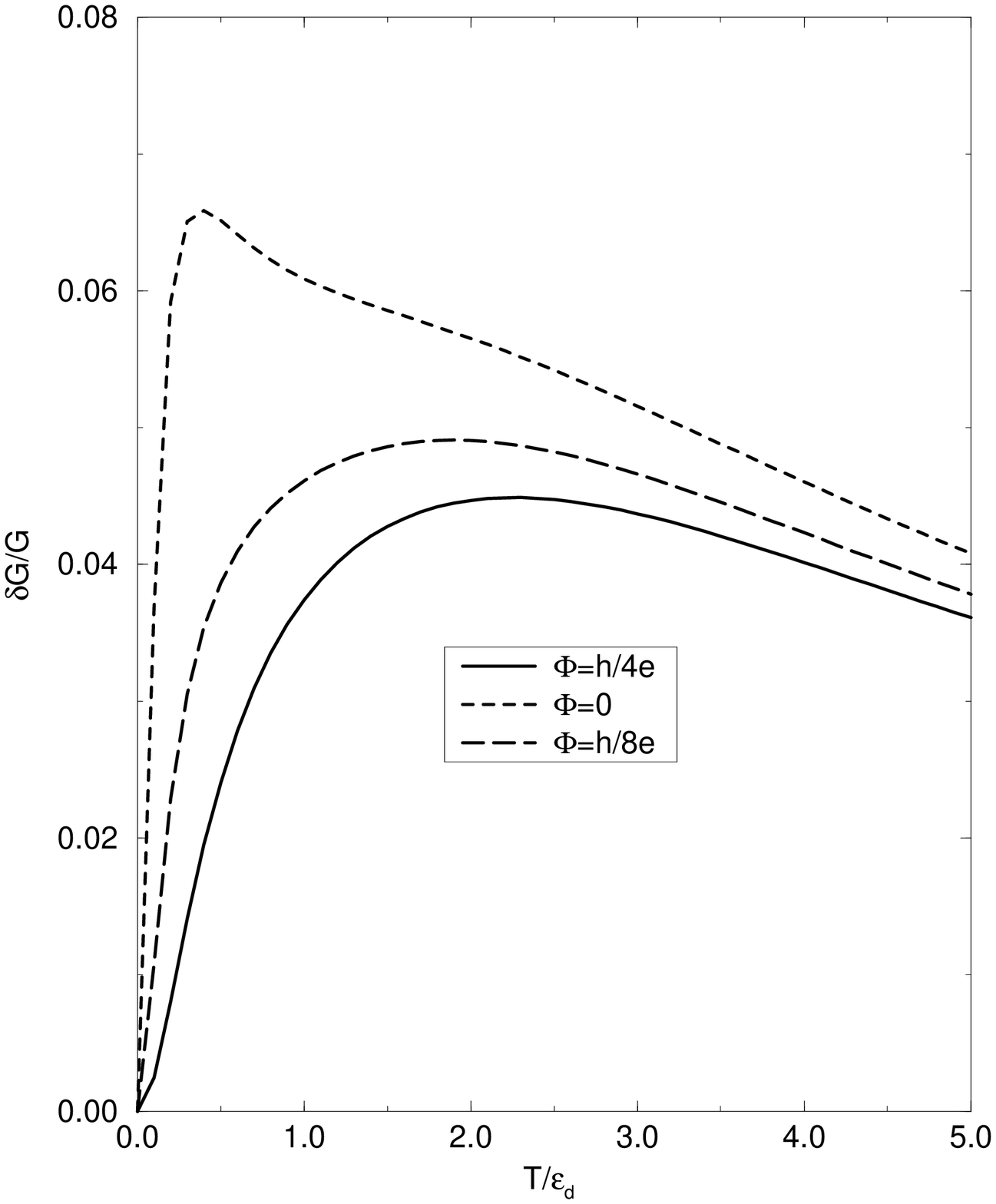}}
    \caption{}
    \end{figure}
  \end{minipage}
  \begin{minipage}[t]{5cm}
    \begin{figure}
    \centerline{\epsfxsize5cm \epsffile{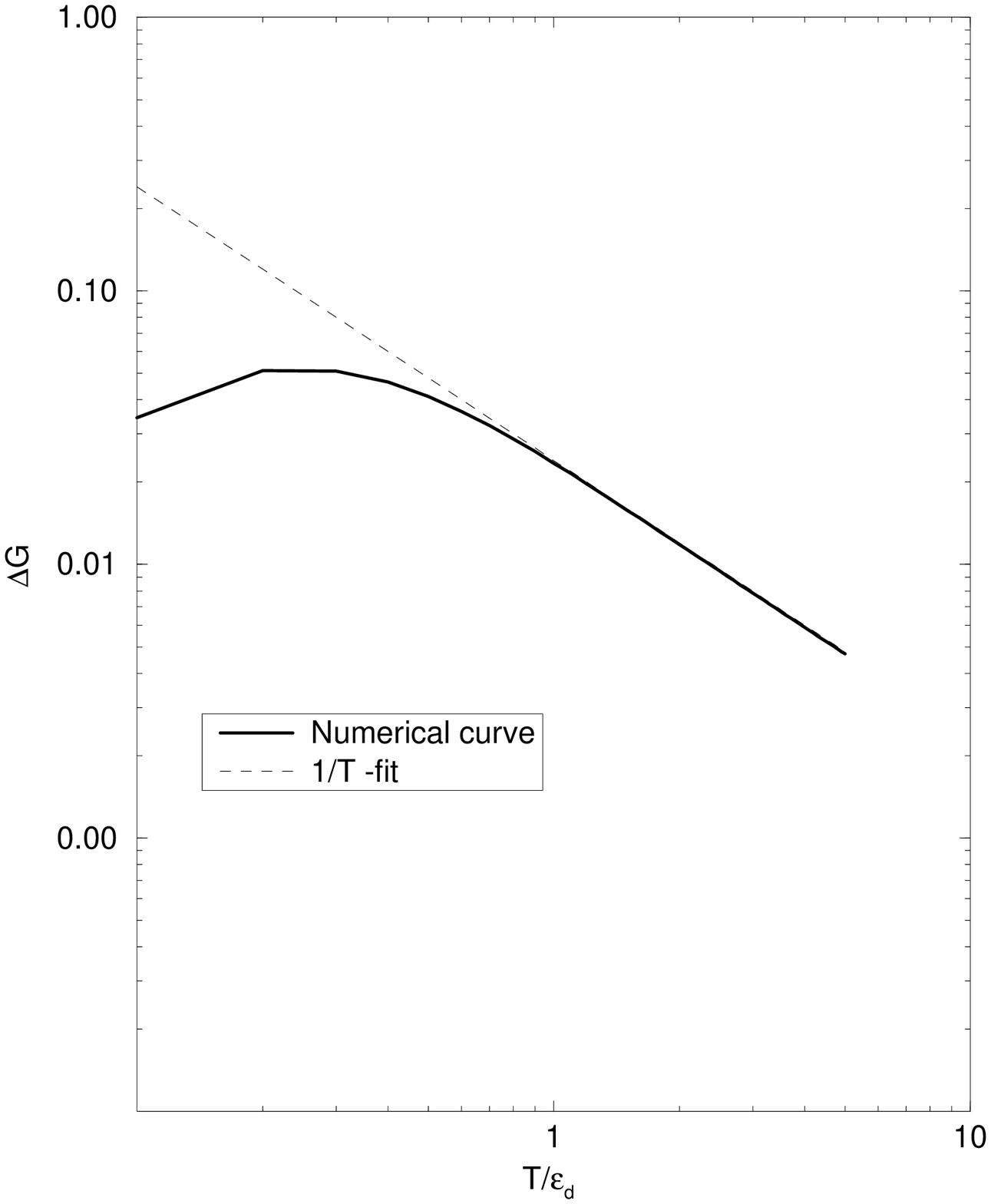}}
    \caption{}
    \end{figure}
  \end{minipage}
\end{minipage}\\

\vspace{12pt}

We acknowledge useful discussions with C.Bruder, H.Courtois, D.Esteve,
B.Pannetier, V.T.Petrashov, G.Sch\"on, B.Spivak and A.F.Volkov. 

\begin{center}
{\large\bf References}
\end{center}

\noindent
{[1]} V.T. Petrashov et al., Phys. Rev. Lett. {\bf 74}, 5268 (1995).\\
{[2]} H. Courtois et al., Phys. Rev. Lett {\bf 76}, 130 (1996).\\
{[3]} A.A. Golubov, F. Wilhelm and A.D. Zaikin, preprint
cond-mat/9601085; submitted to Phys. Rev. B.\\
{[4]} A.F. Volkov, A.V. Zaitsev and T.M. Klapwijk, Physica C {\bf
210}, 21 (1993) and references therein.\\
{[5]} Yu. V. Nazarov and T.H. Stoof, Phys. Rev. Lett. {\bf 76}, 823
(1996); preprint cond-mat/9601071.\\
{[6]} A.V. Zaitsev, Physica B {\bf 203}, 274 (1994).\\

\begin{center}
{\bf Transport coh\'{e}rent dans un fil diffusif entre deux r\'{e}servoirs} 
\end{center}

{\small Nous d\'{e}veloppons une analyse d\'{e}taill\'{e}e du transport
des \'{e}lectrons dans des conducteurs normaux en pr\'{e}sence de
supraconductivit\'{e} induite par l'effet de proximit\'{e}. On a pu trouver
et expliquer la d\'{e}pendance riche
de la conductance, de la densit\'{e} d'\'{e}tat et d'autres
grandeurs en fonction de la temp\'{e}rature et de l'\'{e}nergie. Si le conducteur normal contient une boucle, sa
conductance oscille avec la p\'{e}riode $h/2e$ en fonction du flux
magn\'{e}tique \`{a} travers la boucle. L'amplitude de ces oscillations
en fonction de la temp\'{e}rature fait appara\^{\i}tre un ph\'{e}nomen ``r\'{e}-entrant'' et diminue comme $1/T$ pour $T$ grand.}

\end{document}